\documentstyle[epsfig,12pt]{article}

\catcode`\@=11
\long\def\@makefntext#1{ 
\protect\noindent \hbox to 3.2pt {\hskip-.9pt
$^{{\ninerm\@thefnmark}}$\hfil}#1\hfill} 

\def\thefootnote{\fnsymbol{footnote}}
 \def\@makefnmark{\hbox to 0pt{$^{\@thefnmark}$\hss}}  

\def\ps@myheadings{\let\@mkboth\@gobbletwo
\def\@oddhead{\hbox{} 
\rightmark\hfil\ninerm\thepage}
\def\@oddfoot{}\def\@evenhead{\ninerm\thepage\hfil 
\leftmark\hbox{}}\def\@evenfoot{}
\def\sectionmark##1{}\def\subsectionmark##1{}}

\begin{document}

\newcommand{\symbolfootnote}{\renewcommand{\thefootnote}
        {\fnsymbol{footnote}}}
\renewcommand{\thefootnote}{\fnsymbol{footnote}}
\newcommand{\alphfootnote}
        {\setcounter{footnote}{0}
         \renewcommand{\thefootnote}{\sevenrm\alph{footnote}}}

\newcounter{sectionc}\newcounter{subsectionc}\newcounter{subsubsectionc}
\renewcommand{\section}[1] {\vspace{0.6cm}\addtocounter{sectionc}{1}
\setcounter{subsectionc}{0}\setcounter{subsubsectionc}{0}\noindent
        {\bf\thesectionc. #1}\par\vspace{0.4cm}}
\renewcommand{\subsection}[1] {\vspace{0.6cm}\addtocounter{subsectionc}{1}
        \setcounter{subsubsectionc}{0}\noindent
        {\it\thesectionc.\thesubsectionc. #1}\par\vspace{0.4cm}}
\renewcommand{\subsubsection}[1]
{\vspace{0.6cm}\addtocounter{subsubsectionc}{1}
        \noindent {\rm\thesectionc.\thesubsectionc.\thesubsubsectionc.
        #1}\par\vspace{0.4cm}}
\newcommand{\nonumsection}[1] {\vspace{0.6cm}\noindent{\bf #1}
        \par\vspace{0.4cm}}

\newcounter{appendixc}
\newcounter{subappendixc}[appendixc]
\newcounter{subsubappendixc}[subappendixc]
\renewcommand{\thesubappendixc}{\Alph{appendixc}.\arabic{subappendixc}}
\renewcommand{\thesubsubappendixc}
        {\Alph{appendixc}.\arabic{subappendixc}.\arabic{subsubappendixc}}

\renewcommand{\appendix}[1] {\vspace{0.6cm}
        \refstepcounter{appendixc}
        \setcounter{figure}{0}
        \setcounter{table}{0}
        \setcounter{equation}{0}
        \renewcommand{\thefigure}{\Alph{appendixc}.\arabic{figure}}
        \renewcommand{\thetable}{\Alph{appendixc}.\arabic{table}}
        \renewcommand{\theappendixc}{\Alph{appendixc}}
        \renewcommand{\theequation}{\Alph{appendixc}.\arabic{equation}}
        \noindent{\bf Appendix \theappendixc #1}\par\vspace{0.4cm}}
\newcommand{\subappendix}[1] {\vspace{0.6cm}
        \refstepcounter{subappendixc}
        \noindent{\bf Appendix \thesubappendixc. #1}\par\vspace{0.4cm}}
\newcommand{\subsubappendix}[1] {\vspace{0.6cm}
        \refstepcounter{subsubappendixc}
        \noindent{\it Appendix \thesubsubappendixc. #1}
        \par\vspace{0.4cm}}

\def\abstracts#1{{
        \centering{\begin{minipage}{30pc}\tenrm\baselineskip=12pt\noindent
        \centerline{\tenrm ABSTRACT}\vspace{0.3cm}
        \parindent=0pt #1
        \end{minipage} }\par}}

\newcommand{\bibit}{\it}
\newcommand{\bibbf}{\bf}
\renewenvironment{thebibliography}[1]
        {\begin{list}{\arabic{enumi}.}
        {\usecounter{enumi}\setlength{\parsep}{0pt}
\setlength{\leftmargin 1.25cm}{\rightmargin 0pt}
         \setlength{\itemsep}{0pt} \settowidth
        {\labelwidth}{#1.}\sloppy}}{\end{list}}

\topsep=0in\parsep=0in\itemsep=0in
\parindent=1.5pc

\newcounter{itemlistc}
\newcounter{romanlistc}
\newcounter{alphlistc}
\newcounter{arabiclistc}
\newenvironment{itemlist}
        {\setcounter{itemlistc}{0}
         \begin{list}{$\bullet$}
        {\usecounter{itemlistc}
         \setlength{\parsep}{0pt}
         \setlength{\itemsep}{0pt}}}{\end{list}}

\newenvironment{romanlist}
        {\setcounter{romanlistc}{0}
         \begin{list}{$($\roman{romanlistc}$)$}
        {\usecounter{romanlistc}
         \setlength{\parsep}{0pt}
         \setlength{\itemsep}{0pt}}}{\end{list}}

\newenvironment{alphlist}
        {\setcounter{alphlistc}{0}
         \begin{list}{$($\alph{alphlistc}$)$}
        {\usecounter{alphlistc}
         \setlength{\parsep}{0pt}
         \setlength{\itemsep}{0pt}}}{\end{list}}

\newenvironment{arabiclist}
        {\setcounter{arabiclistc}{0}
         \begin{list}{\arabic{arabiclistc}}
        {\usecounter{arabiclistc}
         \setlength{\parsep}{0pt}
         \setlength{\itemsep}{0pt}}}{\end{list}}

\newcommand{\fcaption}[1]{
        \refstepcounter{figure}
        \setbox\@tempboxa = \hbox{\tenrm Fig.~\thefigure. #1}
        \ifdim \wd\@tempboxa > 6in
           {\begin{center}
        \parbox{6in}{\tenrm\baselineskip=12pt Fig.~\thefigure. #1 }
            \end{center}}
        \else
             {\begin{center}
             {\tenrm Fig.~\thefigure. #1}
              \end{center}}
        \fi}

\newcommand{\tcaption}[1]{
        \refstepcounter{table}
        \setbox\@tempboxa = \hbox{\tenrm Table~\thetable. #1}
        \ifdim \wd\@tempboxa > 6in
           {\begin{center}
        \parbox{6in}{\tenrm\baselineskip=12pt Table~\thetable. #1 }
            \end{center}}
        \else
             {\begin{center}
             {\tenrm Table~\thetable. #1}
              \end{center}}
        \fi}

\def\@citex[#1]#2{\if@filesw\immediate\write\@auxout
        {\string\citation{#2}}\fi
\def\@citea{}\@cite{\@for\@citeb:=#2\do
        {\@citea\def\@citea{,}\@ifundefined
        {b@\@citeb}{{\bf ?}\@warning
        {Citation `\@citeb' on page \thepage \space undefined}}
        {\csname b@\@citeb\endcsname}}}{#1}}

\newif\if@cghi
\def\cite{\@cghitrue\@ifnextchar [{\@tempswatrue
        \@citex}{\@tempswafalse\@citex[]}}
\def\citelow{\@cghifalse\@ifnextchar [{\@tempswatrue
        \@citex}{\@tempswafalse\@citex[]}}
\def\@cite#1#2{{$\null^{#1}$\if@tempswa\typeout
        {IJCGA warning: optional citation argument
        ignored: `#2'} \fi}}
\newcommand{\citeup}{\cite}

\def\fnm#1{$^{\mbox{\scriptsize #1}}$}
\def\fnt#1#2{\footnotetext{\kern-.3em
        {$^{\mbox{\sevenrm #1}}$}{#2}}}

\font\twelvebf=cmbx10 scaled\magstep 1
\font\twelverm=cmr10 scaled\magstep 1
\font\twelveit=cmti10 scaled\magstep 1
\font\elevenbfit=cmbxti10 scaled\magstephalf
\font\elevenbf=cmbx10 scaled\magstephalf
\font\elevenrm=cmr10 scaled\magstephalf
\font\elevenit=cmti10 scaled\magstephalf
\font\bfit=cmbxti10
\font\tenbf=cmbx10
\font\tenrm=cmr10
\font\tenit=cmti10
\font\ninebf=cmbx9
\font\ninerm=cmr9
\font\nineit=cmti9
\font\eightbf=cmbx8
\font\eightrm=cmr8
\font\eightit=cmti8






\def\ms{{\overline{MS}}}
\def\z0{\rm Z^0}
\def\p{\rm p}
\newcommand{\as}{\alpha_{\rm s}}
\newcommand{\epem}{\rm e^+\rm e^-}
\def\pbar{\rm \bar{p}}
\def\mz0{M_{\z0}}
\def\amz{\as(\mz0)}
\def\q{\rm q}
\def\qbar{\rm \bar{q}}
\def\p{\rm p}
\def\pbar{\rm \bar{p}}
\def\n{\rm N}
\def\l{\cal l}
\def\g{\rm g}
\def\x{\rm X}
%
\def\b{\beta }
\def\a{\alpha }
\def\d{\delta }
\def\L{\Lambda }
\def\ra{\rightarrow }
\def\h{\frac{1}{2}}
\topmargin -0.5cm
\oddsidemargin -0.3cm
\evensidemargin -0.8cm
\pagestyle{empty}
\begin{flushright}
{CERN-TH.7465/94}\\
{hep-ph/9410308}
\end{flushright}
\vspace*{5mm}
\begin{center}
{\bf RADIATIVE    QCD CORRECTIONS:}\\{\bf A   PERSONAL OUTLOOK}\\
\vspace*{1cm}
{\bf Andrei L. Kataev}\\
\vspace{0.3cm}
Theoretical Physics Division, CERN,
CH-1211 Geneva 23, Switzerland; \footnote{ On leave of absence from
the
Institute for Nuclear Research of the Russian Academy of Sciences,\\
117312 Moscow, Russia.}\\

\vspace*{1cm}
{\bf ABSTRACT} \\
\end{center}
\vspace*{2mm}
\noindent
We describe several problems related to the studies of the effects
of  radiative QCD corrections in the phenomenological and
theoretical considerations thus
summarizing  the work of the QCD
part of the Symposium on ``Radiative Corrections: Status and
Outlook''.

\vspace*{6cm}
\noindent{ Invited Summary Talk at the Symposium on
``Radiative  Corrections: Status and Outlook'', Gatlinburg,
Tennessee, USA,
27 June -- 1 July  1994}\\

\vspace*{1cm}
\begin{flushleft}
CERN-TH.7465/94\\
hep-ph/9410308\\
October   1994
\end{flushleft}


\newpage


\centerline{\tenbf RADIATIVE QCD CORRECTIONS:
 A PERSONAL OUTLOOK}
\vspace{0.8cm}

\centerline{\tenrm
    ANDREI L. KATAEV}
\baselineskip=13pt
\centerline{\tenit
Theory Division, CERN, CH-1211 Geneva 23, Switzerland\footnote{On
leave of absence from the
 Institute for Nuclear Research of the Russian Academy of Sciences,
Moscow 117312, Russia}}
\vspace{0.9cm}
\abstracts{We describe several problems related to the
studies of the effects of  radiative QCD corrections in the
phenomenological and theoretical considerations thus
summarizing the
work of the QCD part of the Symposium on ``Radiative Corrections:
Status and Outlook''.}
\vspace{1cm}

\section{Introduction}
Some time ago, with other members of the INR group, when we were
completing our joined work \cite{join} on the analytical evaluation
of the high-order QCD corrections to the physical quantity, using the
already developed methods
 \cite{ChT} and the symbolic manipulation
program SCHOONSCHIP \cite{SCH}, we were frequently asking ourselves
 questions about the possible outcome of our long-term project,
which resulted later on in the completing of the calculations of
the next-to-leading-order
 (NLO) QCD corrections to the several deep-inelastic
scattering (DIS) sum rules \cite{sr1,sr2} and of the next-to-next-to-
leading-order (NNLO) QCD corrections to $R(s)=\sigma_{tot}(e^+e^-
\rightarrow\mbox{hadrons})/\sigma_{tot}(e^+e^-\rightarrow\mu^+\mu^-)$
\cite{our}, \cite{ss}. Today the main outcomes of these studies are
more clear to the whole  scientific comunity.

First of all it is important to stress that any cumbersome calculation
of the effects of radiative QCD corrections represents the classical
example of a ``theoretical experiment''. Therefore, as it happens
sometimes with a real phenomenological experiment, the experience gained
after its completon might result in arriving to results that were
not expected at the
beginning of the project. Of course, the most obvious aim of
a precise QCD calculation is related to the phenomenology. Indeed, the
work on the extraction of the values of various QCD parameters from
the existing experimental data clearly necessitates a detailed
consideration of the perturbative QCD effects. These theoretical
contributions can also be further considered as the QCD background
to the effects of electroweak (EW) physics or to
the possible effects of
new physics, say supersymmetry. As the other argument, which
favours a detailed consideration of the perturbative QCD effects,
today we can quote other theoretical studies, pushed ahead by the
NNLO
QCD calculations of $R(s)$ \cite{our}, \cite{ss}, namely the
rediscovery on the more quantitative level of the interesting
world of renormalons in QCD \cite{Zakh}, \cite{Mul}. These works
demonstrated that in order to understand the role of  certain
non-perturbative effects it is also necessary to consider in more
detail the asymptotic structure of the perturbative QCD predictions
in the deep Euclidean region.

Other examples of the importance of the studies of the perturbative
QCD effects to various physical quantities were given in a number
of talks at this very productive  Symposium. Here we  will summarize the
main results of the work of its QCD part and will add some new
information about several subjects,
important from our point of view,
which slipped away from the scientific part of the Symposium.

\section{Determination of the QCD Parameters}
It is known that the standard Lagrangian of the QCD has the
following form
\begin{equation}
L=-\frac{1}{4}F_{\mu\nu}^aF_{\mu\nu}^a+\sum_{f=1}^6 \,
\overline{q}_f(i\hat D-m_f)q_f  ,
\label{lag}
\end{equation}
where $\hat D=\hat\partial-ig\hat A^a\lambda^a/2$. The parameters
$m_i=m_i(\mu)$ are the current quark masses that depend on the
normalization point.
The traditional methods of the
determination of the masses of
light quarks $m_u$, $m_d$ and $m_s$ are the chiral perturbation
theory and the QCD sum rules approach. Recently the new modern
approach based on the
application of  lattice calculations was also
used to determine the value of $m_s$ \cite{lattice}. The masses
of heavy quarks, namely $m_c$ and $m_b$, can be determined using
the QCD sum rules and the methods provided by the potential
models. The theoretical determination of the top-quark mass is
rather   delicate subject. One of the approaches is based on
the study of the fixed points in the solution of the
renormalization-group (RG) equations for the Yukawa couplings in
the general renormalized theory. And of course one can use
experimental data of Fermilab or LEP machines to get phenomenologically
motivated information about the pole mass of the top-quark.
This subject is one of the hottest at present. However, we will
not concentrate on it in this talk. Its more detailed discussion was
presented at this Symposium by Altarelli \cite{Alt}. Here we will make
several comments about the extractions of the masses of lighter quarks.

Let us first mention that the chiral perturbation theory, which
gives the possibility to determine values of the ratios of
light-quark masses that are independent of
  the normalization point
 (for a
recent review see \cite{Leutw}), does not allow us
to fix the scales
at which the corresponding absolute values of the current quark
masses $m_u=4$ MeV, $m_d=7$ MeV and $m_s=140$-$150$ MeV are
defined. It is possible to solve this problem after application
of the finite-energy QCD sum rules technique (FESR) \cite{FESR}
to the two-point function
\begin{equation}
\Pi(q^2)=i\int e^{iqx}\langle TJ(x)J(0)\rangle_0 d^4x
\label{pi}
\end{equation}
of the (pseudo)scalar quark currents $J=m_j\overline{q}_j(\gamma_5)q_j$.
In the  application of this technique, the values of the light-quark
masses explicitly depend on the local duality interval,
namely $m_j=m_j(s_0)$. The transformation to another normalization
point can be made using the solution of the RG
equation:
\begin{equation}
-\frac{\partial\ln m(\mu)}{\partial \ln (\mu^2)}=\gamma_m(\alpha_s)
=\sum_{i\geq 0} \, \gamma_i \bigg(\frac{\alpha_s}{\pi}\bigg)^{i+1},
\label{anom}
\end{equation}
\newpage
where the perturbative expression for the mass anomalous dimension
function is known at the three-loop level \cite{Tar}.

Usually the FESR values of the light-quark masses \cite{GKL},
\cite{Nar1},
\cite{DR} are normalized at \\
1 GeV. The commonly used values
are $m_u$(1 GeV)=6 MeV, $m_d$(1 GeV)=10 MeV, $m_s$(1 GeV)
=$195 \pm 33$ MeV \cite{DR}, which are in agreement with other
determinations \cite{GKL}, \cite{Nar1}. However, it was shown
 \cite{alike}, using the Borel sum rules method \cite{Borel},
that at the scale of over 1 GeV it is important to take into account
the sizeable
instanton-type contributions to the two-point function
of the (pseudo)scalar quark currents. The more concrete calculations
\cite{GN} gave even the explicit form of these instanton contributions
to the theoretical part of the FESR
\begin{equation}
\int_0^{s_0} \rho^{th}(s)ds=\frac{3}{4\pi^2}m(s_0)^2s_0\bigg(1+
R_{ins}(s_0)+R_{pt}(s_0)\bigg),
\label{inst}
\end{equation}
where $R_{pt}(s_0)$ is the perturbative contribution, which is known
at the three-loop level  \cite{GKLS}, and the
expression for $R_{ins}$ reads \cite{GN}:
\begin{equation}
R_{ins}(s_0)=\frac{11}{4\pi}\frac{\hat m_s}{\sqrt s_0}\bigg(\log
\frac{\sqrt s_0}{\Lambda}\bigg)^{8/9}\bigg(\frac{5.17\Lambda}{\sqrt s_0}
\bigg)^9 .
\label{ins}
\end{equation}
The result of Eq. (\ref{ins}) is comparable with the one-loop expression
at $s_0=13\Lambda^2\approx 2$ GeV$^2$. Therefore it is necessary to
understand in more detail the place of these additional theoretical
contributions in the FESR determinations of light-quark masses. It is
possible that these effects are automatically taken into account in
the recent lattice determinations of $m_s$ \cite{lattice}, since this
non-perturbative method was aimed at the calculation   of the $m_s$
value at the high energy scale. The result
obtained in Ref. \cite{lattice}
reads $m_s$(2 GeV)=$127\pm 18$ MeV. Evolving it to the traditional
normalization point using the solution of the RG
equation of Eq. (3) at the three-loop level we get
$m_s$(1 GeV)$\approx 180\pm 30$ MeV, which is in  very good
agreement with the recent determination of the $s$-quark mass from
the Borel sum rules for the divergence of the vector current,
namely $m_s$(1 GeV)=$189\pm 32$ MeV \cite{Jamin}, the latter
being in its turn in agreement with  FESR results
 \cite{GKL},\cite{Nar1},\cite{DR}.
This fact gives us the idea that apart from
 the instanton-type contributions to the correlator of the
(pseudo)scalar quark currents, other uncertainties in the determinations
of the $s$-quark mass are well under  control.

A few
words should be added to the discussions of the heavy-quark
properties, presented at the Symposium in the  theoretical
\cite{Wise},\cite{Fleisch},\cite{Kniehl} and
experimental \cite{Clare}, \cite{Sang}
talks. First of all, it is important to mention that the pole masses
of the heavy quarks are related to the running quark masses by the
following numerical expression \cite{David}
\begin{equation}
m_q(m_q^p) = m_q^p\bigg[1-\frac{4}{3}\frac{\alpha_s(m_q^p)}{\pi}
-\bigg(
14.33-1.04 \sum_{i=u}^{q-1} \, \bigg(
1-\frac{4}{3}\frac{m_i^p}{m_q^p} \bigg)\bigg)
 \bigg(\frac{\alpha_s(m_q^p)}{\pi}\bigg)^2\bigg],
\label{pole}
\end{equation}
where $m_q^p$ are the pole  masses and the coefficient of the
$m_i^p/m_q^p$ term is the approximation, which should be
used only up to the value of the ratio up to 0.3. Traditionally
Eq. (\ref{pole}) was used to determine the values of the running
quark masses from the values of $m_q^p$. However, it was discovered
recently \cite{mren} that the resummation of the
 renormalon-type contributions, non-considered previously,
  to the relations
between heavy-quark running and pole masses result in the appearance
of an additional  contribution, which has the following
approximate form:
\begin{equation}
m_q(m_q^p)\approx m_q^p - \frac{2}{3\beta_0}\Lambda_{QCD}
\label{massren}
\end{equation}
where $\beta_0$ is the first coefficient of the QCD $\beta$-function,
defined as $\beta(\alpha_s)=\mu^2\partial a/\partial \mu^2$
($a=\alpha_s/\pi$),
and $\Lambda_{QCD}$ is the QCD scale parameter in some non-fixed scheme.
Therefore, the pole mass is not defined to an accuracy better than
$\Lambda_{QCD}$ within perturbation theory \cite{mren}. The indirect
indication of this problem comes from detailed studies
 \cite{KK} of the effects of the QCD corrections to
$\Gamma(H^0\rightarrow b\overline{b})=\Gamma_{Hb\overline{b}}$
\cite{GKLS} in the case when the value of $m_b^p$ is considered as
the input parameter and the RG-imroved version of the relation of
Eq. (\ref{pole}) is used to sum up the RG- controllable
$\log(M_H/m_b^p)$-terms and to present the final expression for
$\Gamma_{Hb\overline{b}}$ in terms of $m_b(M_H)$.

Indeed, it was observed  \cite{KK} that
for  reasonably large values of the Higgs boson mass
$M_H$ the corresponding
NNLO corrections to the ratio $R_{Hb\overline{b}}=
\Gamma_{Hb\overline{b}}/\Gamma_{0}^{(b)}$ (where $\Gamma_{0}^{(b)}=
3\sqrt{2}/(8\pi)G_F M_H (m_b^p)^2)$ are larger than
the NLO ones. This observation was further considered  as
an  indication of the asymptotic explosion of the corresponding NNLO
approximation, which is related to the one of Eq. (\ref{pole}). The
latter feature might indicate the importance of the careful treating
of the renormalon contributions to Eq. (\ref{pole}) responsible for
the asymptotic structure of the corresponding perturbative relation.
Now we can convince ourselves that in order to avoid the problem
of the study of the asymptotic behaviour of the perturbative series
related to Eq. (6)
 it might be better to use, in the
phenomenological considerations,
the value of $m_b(\mu\approx M_{\Upsilon}/2
\approx m_b^p)$, evolve it using the RG equation to any high-energy
scale, and then to study the scheme dependence of the corresponding
results. This point of view is in agreement with the one of
Marciano \cite{Marciano} and with the studies of the dependence of
the results of calculations of the QCD corrections to the
$\rho$-parameter on the definition of the top-quark mass
\cite{Sirlin}.
An  analogous
observation came previously from the results of calculations
of the $O(m_b^2/M_Z^2)$
corrections to $\Gamma(Z^0\rightarrow\mbox{hadrons})$
 \cite{ChKK}. Note that a definite attempt to estimate
 the renormalon-type
contributions in Eq. (\ref{massren}) was recently made
\cite{Nar2}. This analysis was based on the following values of
the $c$-quark and $b$-quark running masses in the $\overline{MS}$
scheme: $m_c(m_c^p)=1.23^{+0.02}_{-0.04}\pm 0.03$ GeV,
$m_b(m_b^p)=4.23^{+0.03}_{-0.04}\pm 0.02$ GeV.
The numerical values of the
renormalon contributions $\Delta m_q$ to Eq. (7) were
estimated to be $\Delta m_c \approx 30\pm 20$ MeV and
$\Delta m_b \approx 70$ MeV \cite{Nar2}.
However, we
think that this phenomenological
analysis   is only the first
step in the direction of  future, more definite studies of the
values of these newly discovered effects.

Let us now turn to more classical problems.
The precise determination of the value of the
coupling constant $\alpha_s(\mu)=g(\mu)^2/(4\pi)$ and of the
QCD scale parameter $\Lambda_{\overline{MS}}$ in the $\overline{MS}$
scheme is considered at present as one of the basic phenomenological
tests of QCD.
The detailed update of the determinations of $\alpha_s$ values
from different processes was presented at the Symposium by Bethke
\cite{Bethke}. The results presented are  included in
 Table 1, taken from a later review \cite{Bethke2}.
\newpage
\begin{center}
\begin{tabular}{|l|c|l|l|c c|c|}
   \hline
 &  Q & & &  \multicolumn{2}{c|}
{$\Delta \amz $} &  \\ 
 Process & [GeV] & $\alpha_s(Q)$ &
  $ \amz$ & exp. & theor. & Theory \\
\hline \hline \normalsize
 & & & & & & \\
 DIS [e,$\mu$; Bj-SR] & 1.58
  & $0.375\ ^{+\ 0.062}_{-\ 0.081}$ & $0.122\ ^{+\ 0.005}_{-\ 0.009}
  $ &
  -- & --$$ & NNLO \\
 DIS [$\nu$; GLS-SR] & 1.73
  & $0.32\pm 0.05$ & $0.115\pm 0.006$ & $ 0.005 $ & $ 0.003$ & NNLO \\
 & & & & & & \\
 $R_{\tau}$ [LEP]
  & 1.78 & $0.360 \pm 0.040$ & $0.122 \pm 0.005$
  & 0.002 &  0.004 & NNLO \\
 & & & & & & \\
 DIS [$\nu$; ${\rm F_2\ and\ F_3}$]  & 5.0
  & $0.193\ ^{+\ 0.019\ }_{-\ 0.018\ }$
   & $0.111\pm 0.006$   &
    $ 0.004 $ & $ 0.004$ & NLO \\
 DIS [$\mu$; ${\rm F_2}$]
     & 7.1 & $0.180 \pm 0.014$ & $0.113 \pm 0.005$ & $ 0.003$ &
     $ 0.004$ & NLO \\
& & & & & & \\
 ${\rm Q\overline{Q}}$ states
     & 5.0 & $0.188 \pm 0.018$ & $0.110 \pm 0.006 $ & 0.000 & 0.006
     & LGT \\
 $J/\Psi + \Upsilon$ decays
     & 10.0 & $0.167\ ^{+\ 0.015\ }_{-\ 0.011\ }$ & $0.113\ ^{+\ 0.007\ }
     _{-\ 0.005\ }$ & 0.001 & $^{+\ 0.007}_{-\ 0.005}$ & NLO \\
 & & & & & & \\
 $\epem$ [$\sigma_{\rm had}$]  & 34.0 &
 $0.146\ ^{+\ 0.031}_{-\ 0.026}$ &
   $0.124\ ^{+\ 0.021}_{-\ 0.019}$ & $^{+\ 0.021}_{-\ 0.019}
   $ & -- & NNLO \\
 $\epem$ [ev. shapes]  & 35.0 & \ $0.14\pm 0.02$ &
   $0.119 \pm 0.014$ & -- & -- & NLO \\
 $\epem$ [ev. shapes]  & 58.0 & $0.132\pm 0.008$ &
   $0.123 \pm 0.007$ & 0.003 & 0.007 & resum \\
 & & & & & & \\
 $\p\bar{\p} \rightarrow {\rm b\bar{b}X}$
    & 20.0 & $0.138\ ^{+\ 0.028\ }_{-\ 0.019\ }$ & $0.109\ ^{+\ 0.016\ }
   _{-\ 0.012\ }$ & $^{+\ 0.012}_{-\ 0.007}$ & $^{+\ 0.011}_{-\ 0.010}$ & NLO
\\
 ${\rm p\bar{p},\ pp \rightarrow \gamma X}$  & 24.2 & $0.137
 \ ^{+\ 0.017}_{-\ 0.014}$ &
  $0.112\ ^{+\ 0.012\ }_{-\ 0.008\ }$ & 0.006 &
  $^{+\ 0.010}_{-\ 0.005}$ & NLO \\
 ${\rm p\bar{p} \rightarrow W\ jets}$  & 80.6 & $0.123 \pm 0.025$ &
  $0.121\pm 0.024$ & 0.017 & 0.016 & NLO \\
 & & & & & & \\
$\epem \rightarrow \z0$:  & & & & & & \\
 \ \ $\Gamma (\z0 \rightarrow {\rm had.})$
    & 91.2 & $0.126\pm 0.007$ &
$0.126\pm 0.007$ &
   $ 0.006$ & $^{+\ 0.003}_{-\ 0.004}$ & NNLO \\
 \ \ had. event shapes &
    91.2 & $0.119 \pm 0.006$ & $0.119 \pm 0.006$ &$ 0.001$ & $ 0.006$
& NLO\\ 
 \ \ had. event shapes &
    91.2 & $0.123 \pm 0.006$ & $0.123 \pm 0.006$ & $ 0.001$ & $
0.006$ & resum. \\
 & & & & & & \\
\hline
\end{tabular}
\end{center}
\baselineskip=11.0pt
{\small      \noindent
{\bf Table 1.}
World summary of measurements of $\as$.
Abbreviations:
DIS = deep-inelastic
scattering; GLS-SR = Gross-Llewellyn-Smith sum rules;
Bj-SR = Bjorken sum rules;
LGT = lattice gauge theory;
resum. = resummed NLO.
Reference: S. Bethke, {\it Proc. of the QCD'94},
Montpellier, France, July, 1994;
Aachen preprint PITHA-94-30. The comments on the first result
and its uncertainties are  presented in Section 4.}

Of course, the application of the $\overline{MS}$
scheme represents the example of the phenomenological convention
between theoreticians and experimentalists. Indeed, it does not allow
us to avoid the
theoretical ambiguities due to the existence of the
scale-scheme dependence problem, reviewed at the Symposium by Brodsky
\cite{Brodsky}. However, it is known that in QED the following
relation takes place: $\alpha_{\overline{MS}}(m_e)=\alpha_{OS}[1
+O(\alpha_{OS}^2)]$ where $\alpha_{OS}$ is the QED fine structure
constant. Therefore, the NLO on-shell scheme QED phenomenology is
identical to the $\overline{MS}$ scheme one. Moreover, we think
that the $\overline{MS}$ scheme has also other attractive features
and can be really considered as the conventional referential scheme,
which should be used in the studies of the phenomenological QCD
predictions.
The results of Table 1
give the following modern (but current) world average
value \cite{Bethke} of $\alpha_s$ at the $M_Z$ scale:
$\alpha_s(M_Z)=0.117\pm 0.006$,
which corresponds to the  following values of the parameters
$\Lambda_{\overline{MS}}$:
\begin{equation}
\Lambda_{\overline{MS}}^{(5)}=195^{+80}_{-60}\ MeV,
\Lambda_{\overline{MS}}^{(4)}=280^{+95}_{-80}\ MeV,
\Lambda_{\overline{MS}}^{(3)}=380^{+130}_{-90}\ MeV.
\label{la}
\end{equation}
The cited world average for $\alpha_s$
 is in agreement with
$\alpha_s(M_Z)=0.118\pm 0.006$ given in another recent review work
\cite{Alt1}.

 As was also stressed by Bethke \cite{Bethke}, in order to
provide a more reliable estimate of the theoretical uncertainties
in the extractions of $\alpha_s$ from jet cross sections it is
urgently necessary to calculate still unknown order-$O(\alpha_s^3)$
NLO QCD corrections to the characteristics of  4-jet production in
$e^+e^-$--annihilation. The theoretical background lying under this
problem was discussed at the Symposium by Lampe \cite{Lampe}.
The problem of the calculation of the NLO QCD corrections is
also important for the analysis of the jet production in
$\overline{p}p$ collisions. Indeed, in the talk by Lynn
\cite{Lynn} a
 plot of the data of the CDF collaboration,
taken from Ref. \cite{CDF}, was presented.

One can see that the CDF data \cite{CDF} lie $(1.5$-$2.4)\sigma$
below the
 range of the QCD predictions. Clearly this fact stimulates
further experimental and theoretical studies. The status of the
theoretical program of  the jet cross section calculations in
the $\overline{p}p$ collisions was discussed by Giele \cite{Giele}. It
was emphasized that the interest in the corresponding NLO calculations
is stimulated by the fact that the high-multiplicity jet cross sections
are only known at the leading-order (LO) level. One of the most
important messages given by Giele \cite{Giele} is that the continuous
development of the theoretical technology \cite{BernDixKos},\cite{Kunszt}
can result in the appearance
of the NLO calculations to $\overline{p}p\rightarrow 3$ jets and
$\overline{p}p\rightarrow W,Z+2$ jets cross sections before
 the end
of this year.

The interesting phenomenological result, related to the measurement
of the
3-jet cross sections by the CLEO group from Cornell, was presented
at the Symposium by Sanghera \cite{Sang}. These measurements allowed
one to determine the value of $\alpha_s$
 in the energy region with four active
flavours and to obtain the following result
$\alpha_s(10.53$ GeV)=0.164 $\pm 0.004$(exp)$ \pm 0.015$ (theory). The
RG evolution of this result
through the threshold of the production of the $b$-quark gave
the following value :
$\alpha_s(M_Z)=0.113 \pm 0.002 \pm 0.007$
\cite{Sang}. This results is
in reasonable agreement with the world average
value of $\alpha_s$. Note, however, that the measurement
of $\alpha_s$ from the jets rates directly at a $Z^0$-pole
give the larger central value of $\alpha_s$ (see Table 1).

In general
the determination of $\alpha_s$ from the LEP measurements
creates a definite problem in the comparison with certain
other results, say with the one presented by Sanghera \cite{Sang}.
 Indeed, the central values of the results obtained from LEP data
 are usually higher than the results extracted from
DIS, which are in agreement with the result of the
CLEO analysis.
For example, the most detailed recent extraction of $\alpha_s$ from
the hadronic width of the $Z^0$ using three independently
written computer codes BHM, TOPAZ0 and ZFITTER
gave the following value  \cite{Heb}:
$\alpha_s(M_Z)=0.120\pm 0.007$ (exp) $\pm \delta\alpha_s$(theor),
where
\begin{equation}
\delta\alpha_s(theor)=\pm 0.002 (EW) \pm 0.002 (QCD)^{+0.004}_{-0.003}
(m_{t}^p,M_H).
\label{uncert}
\end{equation}
 This result, discussed in the talks of
Bethke \cite{Bethke}, Fleischer \cite{Fleisch} and
 Hollik
\cite{Hollik} (who described the current status of the comparisons
of the analyses of the EW$\times$QCD effects at the $Z^0$-pole
by means of the different computer codes within the program of
the LEP 1 Theoretical Working Group \cite{Bardin}
) is in agreement
with the one obtained  with the help of the
LEPTOP program \cite{Leptop}
, namely $\alpha_s(M_Z)=0.125 \pm 0.005$ (exp)
$\pm 0.002$ (theory). Clearly, the results of Refs.
\cite{Heb}, \cite{Leptop} are larger than the extractions
of $\alpha_s$ from the DIS data (see Table 1 and Ref. \cite{ChK}).
Moreover, the corresponding value of $\Lambda_{\overline{MS}}^{(3)}$,
which follows from the analysis of the LEP data,
 lies above  the bound
$\Lambda_{\overline{MS}}^{(3)}\leq 400$ MeV, extracted
 \cite{Grozin} from the analysis of the
low-energy $e^+e^-$ data by means of  the QCD sum rules approach.
 This deviation creates
a problem in the understanding of the relation of the results
 of the QCD sum
rules analysis, which are  known to be quite successful in
 describing the low-energy hadronic phenomenology, to the
  above-mentioned LEP values of $\alpha_s$.
We think that before making any definite conclusions
it is necessary to reconsider the problem of the extraction
of the parameter $\Lambda_{\overline{MS}}^{(3)}$ from the
low-energy $e^+e^-$ data, taking into account the sizeable
 $O(\alpha_s^3)$ perturbative contributions to the Borel sum rules,
which are known from the results of Ref. \cite{jetplet}.

\section{New Theoretical Results}
The work on the precise analysis of the LEP experimental data
stimulates the calculations and estimates of new theoretical
effects. It is clear that large
corrections in the Standard Model can
come in particular from two
sources, namely  from the
top-quark mass-dependent effects and from the effects
of the higher-order perturbative QCD corrections.
 Two new results of the calculation of the
top-quark mass-dependent QCD contributions were actively discussed
in the talks of Hollik \cite{Hollik}, Kniehl \cite{Kniehl}
and Fleischer \cite{Fleisch}. The first important result is the
analytical calculation \cite{Avdeev}
of the three-loop QCD corrections to the
Veltman $\rho$-parameter,
 which is defined as the ratio of the neutral to charged currents
amplitudes:
$\rho=G_{NC}(0)/G_{CC}(0)=1/(1-\Delta\rho)$,
where
\begin{equation}
\Delta \rho \approx 3 x_t (1+\rho^{EW}x_t)(1+\delta^{QCD})
\label{delta}
\end{equation}
$x_t=\sqrt{2}G_{\mu}(m_t^p)^2/(16\pi^2)$ and $\rho^{EW}$
is the two-loop EW contribution, calculated for different
values of $M_H$ in  several works \cite{EWrho}. The
 QCD contribution to Eq. (\ref{delta}),
normalized at the pole top-quark  mass,
has the following numerical form \cite{Avdeev}
\begin{equation}
\delta^{QCD}=-2.86 a(m_t^p)-(21.27-1.79n_f) a(m_t^p)^2 ,
\label{resa}
\end{equation}
where $a(\mu)=\alpha_s(\mu)/\pi$. As was mentioned by Kniehl
\cite{Kniehl} and Fleischer \cite{Fleisch}, it is interesting
to study the numerical effects of this result using different
prescriptions to fixing the renormalization-scheme ambiguities,
namely the effective charges approach (ECH) \cite{ECH}, the principle
of minimal sensitivity (PMS)  \cite{PMS},  the
BLM procedure \cite{BLM} and the different definitions of the
top-quark mass. Even more important is the detailed consideration
of the scheme dependence of the EW contributions to the
$\rho$-parameter \cite{Bochkarev}.

Other recent works \cite{singlet} concentrated on the calculation
of the   $O(a^3)$ contributions to $\Gamma_{Z^0}=
\Gamma(Z^0\rightarrow
\mbox{hadrons})$ from the singlet diagrams  with the loop formed
by the propagaton of the virtual
top-quark. Including the previously calculated similar term of order
$O(a^2)$  \cite{KnKuhn}, and the results of Refs. \cite{our},
\cite{ss},
 one gets the following  $O(a^3)$ QCD approximation
for $\Gamma_{Z^0}$:
\begin{eqnarray}
\Gamma_{Z^0}&=&\Gamma^{QPM}\bigg[\sum_{i=u}^{b} \, (g_i^V)^2
(1+a_{5}+1.409a_{5}^2-12.767a_{5}^3)+
\bigg(\sum_{i=u}^{b} \, g_i^V\bigg)^2
(-0.413a_{5}^3) \nonumber \\
&&+\sum_{i=u}^{b} \, (g_i^A)^2 (1+ a_{(5)}+1.409a_{(5)}^2-12.767
a_{(5)}^3 ) \nonumber \\
&&+
 \bigg[\bigg(-\frac{37}{12}
-\ln\bigg(\frac{\overline{m}_t^2}
{M_Z^2}\bigg)\bigg)a_{(5)}^2  \nonumber \\
&&+\bigg(-18.65+\frac{23}{12}\ln^2\bigg(\frac
{\overline{m}_t^2}{M_Z^2}\bigg)
-\frac{67}{18}\ln\bigg
(\frac{\overline{m}_t^2}{M_Z^2}\bigg)\bigg)a_{(5)}^3\bigg]\bigg]
\label{tot}
\end{eqnarray}
where  $\Gamma^{QPM}=G_FM_Z^3/(8\pi\sqrt{2})$,
$g_i^V=2I_i-4Q_is_W^2$, $g_i^A=2I_i$, and where
$a_{(5)}=a(M_Z)$
corresponds to $f=5$ numbers of flavours and $\overline{m}_t=
m_t(m_t^p)$.
For  simplicity we neglected
the known  $O(m_b^2/M_Z^2)$
corrections \cite{ChKK} and the  $O(\alpha\alpha_s)$ corrections
\cite{Kataev}, which  should be taken into
account in the precise analysis of the LEP data.

In the present
situation, when the LEP experimental data on $\Gamma_{Z^0}$
is continuously increasing, and in view of the existence of the
certain deviation of the central values of $\alpha_s(M_Z)$ (and thus
$\Lambda_{\overline{MS}}$) from the ones, extracted from the DIS
and low-energy hadronic phenomenology, one can ask
whether taking into account  higher-order perturbative
QCD terms can affect the outcomes of the fits of the LEP data. In order
to study this question
 the effects of the
higher-order
QCD corrections were estimated \cite{KatSt1,KatSt2}
using the procedure,
proposed in Ref. \cite{PMS},  of the re-expansion
of the optimized expression for the physical quantities into the initial
$\overline{MS}$ scheme. The results of applications of this
procedure to the Euclidean $D$-function
\begin{equation}
D(Q^{2}) = Q^{2} \int^{\infty}_{0} \, \frac{R(s)}{(s+Q^{2})^{2}} \, ds
= 3 \Sigma Q^{2}_{f} \bigg[1 + a + \sum _{i\geq 1} \,
 d_{i}a^{i+1}  \bigg]
+  (\Sigma Q_{f})^{2} [ \tilde{d}_{2} a^{3} + O(a^{4}) ] ,
\label{13}
\end{equation}
for  different values of the number of flavours $f$, are presented
in Table 2 taken from Ref. \cite{KatSt2}.
\begin{center}
\begin{tabular}{|c|c|c|c|c|} \hline
$f$ & $d^{ex}_{2}$ & $d^{est}_{2}$ &
$d^{est}_{3}$ & $ d_4^{est}-c_3d_1$\\ \hline
1 & 14.11 & 7.54 &  75 & 474 \\ \hline
2 & 10.16 & 6.57 &  50 & 261 \\  \hline
3 & 6.37 & 5.61 &  27.5 & 111 \\ \hline
4 & 2.76 & 4.68 &  8  & 23 \\  \hline
5 & $-$0.69 & 3.77 &  $-$8 & $-$15  \\ \hline
6 & $-$3.96 & 2.88 &  $-$21 & $-$1.8\\ \hline
\end{tabular}
\end{center}
\baselineskip=11.0pt
{\small \noindent
{\bf Table 2.}
 The results of estimates of the $O(a^3)$, $O(a^4)$ and $O(a^5)$
 corrections in
the series for the
$D$-functions.}

The estimated value of the NNLO term $d_2^{est}$ is in
reasonable agreement with the result $d_2^{ex}$ of the exact calculations
\cite{our}, \cite{ss}.

In order to obtain the related estimates of the coefficients of
$R(s)  =  3 \Sigma Q^{2}_{f} [1 + a_{s} + \sum_{i\geq 1}
r_{i} a^{i+1}_{s} ]
 +  (\Sigma Q_{f})^{2} [ \tilde{r}_{2} a^{3}_{s} + ... ]$
it is necessary to take into account the effects of the analytical
continuation to the Minkowskian region. The corresponding
coefficients $r_i$ have the following form:
$r_{1}  =  d_{1}$,
$r_{2} =  d_{2} - \pi^{2} \beta^{2}_{0}/3$,
$\tilde{r}_{2} = \tilde{d}_{2}$,
$r_{3}  =  d_{3} - \pi^{2} \beta^{2}_{0}
(d_{1} + \frac{5}{6} c_{1})$ and
$r_{4} = d_{4} -\pi^2\beta_0^2 (2d_2+\frac{7}{3}c_1d_1
+\frac{1}{2}c_1^2+c_2)+\frac{\pi^4}{5}\beta_0^4$, where
$c_i$ are defined through the coefficients of the QCD $\beta$-function
as $\beta(a)=-\beta_0a^2(1+\sum_{i\geq 1} c_ia^i)$. Using these
expressions, we arrive at the
results of Table 3, taken from the revised version of \\
Ref. \cite{KatSt2}.
\begin{center}
\begin{tabular}{|c|c|c|c|c|} \hline
$f$ & $r^{ex}_{2}$ & $r^{est}_{2}$ &
$r^{est}_{3}$ & $r_4^{est}-c_3d_1$\\ \hline
1 & $-$7.84 & $-$14.41 &  $-$166 & $-$1748\\ \hline
2 & $-$9.04 & $-$12.63 &  $-$147 & $-$1156 \\ \hline
3 & $-$10.27 & $-$11.03 &  $-$128 & $-$669 \\ \hline
4 & $-$11.52 & $-$9.58 &  $-$112 & $-$263 \\ \hline
5 & $-$12.76 & $-$8.29 &  $-$97 & 64 \\ \hline
6 & $-$14.01 & $-$7.17 &  $-$83 & 334 \\ \hline
\end{tabular}
\end{center}
\baselineskip=11.0pt
{\small \noindent
{\bf Table 3.}
The results of estimates of the $O(a^3)$, $O(a^4)$ and $O(a^5)$
 corrections in
the series for $R(s)$.}

It should be stressed that for $\alpha_s(M_Z)\approx 0.12$
the corresponding  $O(a^4)$ contributions
to $\Gamma_{Z^0}$, as estimated in Refs. \cite{KatSt1,KatSt2},
namely
\begin{eqnarray}
\delta\Gamma_{Z^0}&=&\Gamma^{QPM}\bigg[\sum_{u}^{b}(g_i^V)^2
+(g_i^A)^2\bigg](-97a_{(5)}^4) \nonumber \\
&=&\Gamma^{QPM}\bigg[\sum_{u}^{b}(g_i^V)^2
+(g_i^A)^2\bigg](-2\times 10^{-4}),
\label{cor}
\end{eqnarray}
are of the same
order of magnitude as the  $O(m_b^2/M_Z^2)$ corrections,
involved in the current analysis of the LEP data. However, since
they are negative, they can only increase slightly the central value
of $\alpha_s(M_Z)$ and will thus not remove the  deviations
of the values of $\alpha_s(M_Z)$ extracted from different
 processes.
 The positive  $O(a^5)$ corrections to $\Gamma_{Z^0}$, as
estimated in Ref. \cite{KatSt2} (see also the last column of Table 2),
are small and can be safely neglected.

Let us now turn to the analysis of the perturbative predictions
for the ratio $R_{\tau}=\Gamma(\tau\rightarrow\nu_{\tau}+
\mbox{hadrons})/\Gamma(\tau\rightarrow\nu_{\tau}\overline{\nu}_ee)$,
which were not discussed at this Symposium. It is known that this
ratio is related to $R(s)$ by the following physical FESR
\cite{BNP}:
\begin{equation}
R_{\tau}  =  2 \int_{0}^{M^{2}_{\tau}} \, \frac{ds}{M^{2}_{\tau}} \,
(1 - s/M^{2}_{\tau})^{2} \, (1 + 2s/M^{2}_{\tau}) \tilde{R}(s)
 \simeq  3[1 + a_{\tau} + \sum_{i\geq 1} \,
r_{i}^{\tau} a^{i+1}_{\tau} ]
\label{19}
\end{equation}
where $a_{\tau} = \alpha_{s} (M^{2}_{\tau}) / \pi$ and $\tilde{R} (s)$
is $R(s)$
with $f = 3, (\Sigma Q_{f})^{2} = 0, 3 \Sigma Q^{2}_{f}$ substituted
for $3 \Sigma \mid V_{ff'} \mid^{2}$ and $\mid V_{ud} \mid^{2} +
\mid V_{us} \mid^{2} \approx 1$.

It was shown  \cite{pp} that it is convenient to
express the coefficients of the
series (15) through  those  of the series (13) for the
$D$-function in the following form :
$r^{\tau}_{1} =
d_{1}^{\overline{MS}} (f = 3) +  g_1$,
$r^{\tau}_{2} =
d_{2}^{\overline{MS}} (f = 3) +  g_2$,
$r^{\tau}_{3}  =
d_{3}^{\overline{MS}} (f = 3) + g_3$ ,
where the numerical expressions for the coefficients $g_i$ read
$g_1  = 3.563$,
$g_2  = 19.99$,
$g_3  = 78.00$.
One of the pleasant features of these coefficients is that
they are absorbing all effects of the analytical continuation.
Following the lines of Ref. \cite{pp} we derive the corresponding
expression for the coefficient $r^{\tau}_{4}$:
$r^{\tau}_{4} = d_4^{\overline{MS}}(f=3) + g_4$
where
\begin{eqnarray}
g_4&=&-[4d_3+3d_2c_1+2d_1c_2+c_3]\beta_0I_1+[6d_2+7c_1d_1+
\frac{3}{2}c_1^2+3c_2]\beta_0^2I_2   \nonumber \\
&& -[4d_1+\frac{13}{3}c_1]\beta_0^3I_3+
\beta_0^4I_4 = 3.6c_3(f=3)+14.25d_3(f=3)-466.8
\label{g4}
\end{eqnarray}
and $I_4=41041/864-265\pi^2/36+\pi^4/5 \approx -5.668$, while
the expressions for $I_1$, $I_2$ and $I_3$ are known from the
considerations of Ref. \cite{pp}.
Using now the estimates of Table 2
we get the estimates of the
corresponding higher-order coefficients of $R_{\tau}$
\cite{KatSt1,KatSt2}: $(r_3^{\tau})_{est}
\approx 105.5 a_{\tau}^4$,
$(r_{4}^{\tau})_{est}\approx 94 a_{\tau}^5$. It will be interesting
to include these estimates in the analysis of the experimental data.

Some work in this direction  was
recently done in Ref. \cite{Dib}, where
the experimental data of the
ALEPH group for $R_{\tau}$ and their moments
was used to estimate from the fit
the value of the coefficient $d_{3}(f=3)$
of the $D$-function. The result of this fit is presented
in Fig. 2 taken from Ref. \cite{Dib}.

\newpage
In the process of the fits of Ref. \cite{Dib}, it was assumed that
the non-perturbative contributions to $R_{\tau}$ can be neglected.
For $\mu=1$ GeV the result of the fit is $d_3^{est}=29 \pm 4 \pm 2$
\cite{Dib}, where the second uncertainty is due to the theoretical
uncertainties of the  procedure used. The obtained result is in
very good agreement with the results of theoretical considerations
\cite{KatSt1,KatSt2} (see also Table 2). In fact this agreement
is not very surprising, since the estimate of  Ref. \cite{Dib} was
obtained with the help of the ``optimization'' of the experimental
part of the relation $R_{\tau}^{theory}=R_{\tau}^{exp}$, while
the theoretical estimates \cite{KatSt1,KatSt2} were obtained from
the ``optimized'' theoretical l.h.s. of this relation. However, it
is plesant to see that this procedure is respected by the analysis
of the experimental data, presented in Fig. 2.
This fact can be considered as the additional
argument in favour of applicability of the methods of the estimates
of the higher-order perturbative corrections proposed in
Ref. \cite{PMS} and further used in the theoretical considerations
of Ref. \cite{KatSt1,KatSt2}.

\section{Polarized Deep-Inelastic Scattering}
The detailed analysis of  DIS is traditionally considered as
one of the basic problems of the physics of strong interactions.
 Special attention is nowadays paied to the experimental and
theoretical consideration of the polarized DIS processes and especially
of the polarized DIS sum rules (SRs), namely of
the Bjorken polarized SR
\begin{equation}
\Gamma_1^{p}-\Gamma_1^{n}=\int_0^1 \,
\big[g_1^{ep}(x,Q^2)-g_1^{en}(x,Q^2)\big]dx
\label{bjsr}
\end{equation}
and of the Ellis-Jaffe SR
\begin{equation}
\Gamma_1^{p(n)}=\int_0^1 \, g_1^{p(n)}(x,Q^2) dx .
\label{ejsr}
\end{equation}
The current status of the investigations of this very hot topic
was discussed at the Symposium in the beautiful experimental-oriented
talk of Stuart \cite{Stuart} and theoretically-oriented talk of
Forte \cite{Forte}. Clearly, experiment is going ahead of the theory
in this field. Indeed, two new measurements of the proton polarized
structure function $g_1^p(x,Q^2)$ in different energy regions became
recently available. The result of the measurement at the energy
scale $Q^2=10$ GeV$^2$, as presented
at Fig. 3, taken from Ref. \cite{SMC},
was obtained by the SMC group at CERN.
The preliminary data of the E143 group, who  measured
$g_1^{p}$ at $Q^2=3$ GeV$^2$, depicted in Fig. 4,
was presented by Stuart \cite{Stuart}.

After extrapolation of the results of the measurements of the SMC and
E143 groups in the region of small and large $x$ and integration
over the whole region $0\leq x\leq 1$ the following new experimental
results were obtained \cite{SMC}, \cite{E143}:
\begin{eqnarray}
\Gamma_1^{p}(Q^2=10\ GeV^2)&=&0.136 \pm 0.011\pm 0.011 \nonumber \\
\Gamma_1^{p}(Q^2=3\ GeV^2)&=&0.129 \pm 0.004\pm 0.010 .
\label{ejres}
\end{eqnarray}
The latter result is of particular interest since, after combining
it with the previous neutron data of the E142 collaboration
at an average $Q^2$ of 2 GeV$^2$, namely
$\Gamma_1^n=-0.022 \pm 0.011$ \cite{E142}, it
is possible to ``measure'' the value of the polarized Bjorken
SR in the low-energy region. The re-analysis of these data, made
in Ref. \cite{EK1}, resulted in the definite modification of
the original E142 result  \cite{E142}. In fact the
authors of Ref. \cite{EK1} obtained the following number:
\begin{equation}
\Gamma_1^n(Q^2=2\ GeV^2)=-0.028 \pm 0.006 (stat) \pm 0.009 (syst).
\label{ejmod}
\end{equation}
Combining Eq. (\ref{ejmod}) with the first, preliminary, E143 result
for the proton, namely $\Gamma_1^p(Q^2=3\ GeV^2)=0.133\pm0.004\pm0.012$,
the authors of Ref. \cite{EK2}
extracted the value of the Bjorken polarized SR at
$Q^2=2.5$ GeV$^2$:
\begin{equation}
\Gamma_1^p-\Gamma_1^n(Q^2=2.5\ GeV^2)=0.161 \pm 0.007\pm 0.015 .
\label{bjp}
\end{equation}
This result was further used to compare the results
discussed above, which were
experimentally motivated,  with the theoretical QCD
predictions, which we are now going to discuss.

A  number of works are devoted to the calculation of
the theoretical contributions into the polarized Bjorken SR. Summarizing
the available information, let us present the theoretical expression
for this fundamental SR, normalized to $f=3$ numbers of flavours:
\begin{equation}
\Gamma_1^p(Q^2)-\Gamma_1^n(Q^2)=\frac{1}{6}|\frac{g_A}{g_V}|
\bigg[1-a-3.583a^2-20.215a^3 \nonumber \\
-130a^4-O(a^5)\bigg] -O\bigg(\frac{1}{Q^2}\bigg).
\label{bjpth}
\end{equation}
The exact expressions for the  $O(a^2)$ and $O(a^3)$ corrections
are known from the calculations of Refs. \cite{sr2}, \cite{NZ1}
and Ref. \cite{LV} respectively. The $O(a^4)$ coefficient was
estimated in Refs. \cite{KatSt1,KatSt2}, while a
less substantiated
estimate of the $O(a^5)$ term can be read from the results of
the studies of Ref. \cite{KatSt2}. The estimates of the
$O(a^4)$ terms are in qualitative agreement with the results of
applications of the Pad\'e resummation technique \cite{Sam}.

The  expression for the Ellis-Jaffe SR consists
of two parts, namely from the non-singlet and the
singlet contributions.
The  $O(a^2)$ correction to the singlet contribution
was calculated in Ref. \cite{Larin} (see also Ref. \cite{Van2}),
 while the estimates
of the  $O(a^3)$ corrections were  given recently
\cite{Kataev2}. Combining these results with the ones for
the non-singlet contribution (which coincide with the
Bjorken polarized SR) one can arrive at the following
expression for the Ellis-Jaffe SR in the $\overline{MS}$
scheme
\begin{eqnarray}
\Gamma_1^{p(n)} (Q^2)  &=& \bigg[ 1-a-3.583a^2 - 20.215a^3 - 130
a^4\bigg] \times \left( \pm {1\over 12} a_3 + {1\over 36} a_8\right)
\nonumber \\
&& +\bigg[ 1-a-1.096a^2 - 3.7a^3 \bigg] ~{1\over 9} ~
\Delta\Sigma (Q^2)~+O\bigg(\frac{1}{Q^2}\bigg)
\label{20}
\end{eqnarray}
where $a_3=\Delta u-\Delta d$, $a_8=\Delta u+\Delta d-2\Delta s$,
$\Delta\Sigma=\Delta u+\Delta d+\Delta s$ and $\Delta u$, $\Delta d$
and $\Delta s$ can be considered as the measure of polarization of
the quarks in a nucleon.

The results of Eqs. (\ref{bjpth}), (\ref{20}) were
recently used  \cite{EK2} to determine the values of
$\alpha_s$, $\Delta s$ and $\Delta\Sigma$ from the available
experimental data on polarized SRs. At the first stage of the
analysis, the authors of Ref. \cite{EK2} neglected the higher-twist
(HT)
 $O(1/Q^2)$ contributions to Eqs. (\ref{bjpth}),
and (\ref{20}), used the  experimentally motivated
result of Eq. (\ref{bjp}) that they derived
for the polarized Bjorken SR and
extracted the value of $\alpha_s(Q^2=2.5\ GeV^2)$ in different
orders of perturbation theory. The outcomes of this analysis
are depicted at Fig. 5, taken from Ref. \cite{EK2}.
The final result, obtained by taking  account of
the estimates of the $O(a^4)$ corrections given in Refs.
\cite{KatSt1,KatSt2}, is $\alpha_s(2.5$ GeV$^2)=0.375^{+0.062}
_{-0.0081}$ \cite{EK2}. This result corresponds to the
following value of $\alpha_s$ at the $M_Z$-pole: $\alpha_s(M_Z)
=0.122^{+0.005}_{-0.009}$ \cite{EK2}. Using this result
the authors managed further on to adjust  all available
experimental data on the polarized Ellis-Jaffe SR with  the
assumption, that $\Delta s \neq 0$. The outcomes of this analysis,
presented in Fig. 6 taken from Ref. \cite{EK2}, give $\Delta s
=-0.10\pm 0.03$.

And finally, using the facts that the perturbative corrections
to the Ellis-Jaffe SR are negative and that the absolute values
of the perturbative contributions to the non-singlet part
are significantly larger than the ones to the singlet part
(see \\Eq. (\ref{19}), the authors of Ref. \cite{EK2}
extracted the
average value of $\Delta\Sigma(Q^2=10$ GeV$^2$) from the
fits of all the data
available up to now  (see Fig. 7 taken from
Ref. \cite{EK2}).
Notice that Fig. 7 demonstrates that it is possible to
satisfy all available data, including the neutron data
of the E142 collaboration, by the condition $\Delta\Sigma=0.31
\pm 0.07$ only after taking into account higher-order perturbative
corrections.

This conclusion gives  one more example of the
importance of the consideration of  the higher-order
perturbative QCD effects.

In spite of the fact that  the authors
of Ref. \cite{EK2} used in their analysis
very preliminary E143 data, which
 were higher than the final
result of Ref. \cite{E143}
due to the non-careful treatment
 of the high $x$-extrapolation
(for a detailed discussion of this important subject see
the talk by Forte \cite{Forte}),
their results
have  rather solid status. Indeed, they are in agreement
with the results $\Delta s(\infty)=-0.097 \pm 0.018$,
$\Delta\Sigma(\infty)=0.33\pm 0.04$ and
$\alpha_s(2$ GeV$^2$)=$0.49\pm 0.06$, which corresponds to
$\alpha_s(M_Z)=0.125 \pm 0.06$ \cite{AR1}. These results were
obtained from the careful fits of all available data on polarized
DIS in the approximation when HT contributions are neglected
\cite{AR1}. The interesting physics, lying beyond the outcomes
of the fits of Refs. \cite{EK2,AR1}, was discussed at the Symposium
by Forte \cite{Forte} and more recently in the work of Ref. \cite{AR1}.

Let us now summarize  the current status of the studies of the
 HT corrections to DIS SRs. The concrete form
of the matrix elements contributing to the HT corrections
are known from the considerations of
Ref. \cite{VSh}. The numerical values of these matrix elements
were calculated using different approaches. It is known that the
results of the three-point function QCD sum rules calculations
\cite{BBK} are larger than the ones obtained with the help of the
bag model calculations \cite{JU}. The original considerations
 \cite{BBK} were recently reanalysed  \cite{RR}
with the help of a similar approach.
The estimates obtained  in Ref. \cite{RR}  turned out to be
larger than the original results of  \cite{BBK} but they
have surprisingly small error bars. If one adopts the 50$\%$
error bars to the results of Ref. \cite{RR}, which are typical
of the physical outcomes of the three-point function QCD sum
rules calculations, one can convince oneself that within
these error bars the results of Ref. \cite{BBK} remain true.
Moreover, these results  were
recently confirmed
 \cite{Mank} by the outcomes of the
application of the same method to the three-point function with
different interpolating currents. In our opinion, these facts
provide a solid background to the applications of the
results  \cite{BBK} in the phenomenological studies
of the experimental data for
the polarized Bjorken SR. In fact this
was already done in Ref. \cite{EK2}, where it was demonstrated
that the incorporation of the HT corrections into the theoretical
expression for the polarized Bjorken SR gives  a  smaller
value  $\alpha_s(M_Z)=0.118^{+0.007}_{-0.014}$.
This result is in agreement with the extraction of $\alpha_s$
from the Gross-Llewellyn Smith SR,
  taking into account
HT corrections \cite{BrKol}, namely
 $\alpha_s(M_Z)=0.115\pm 0.006 ($exp)
$\pm 0.003$ (theory) \cite{ChK} (see the
second result presented in Table 1).
The second
result  \cite{EK2} is also in agreement with the
value  of $\alpha_s$ obtained from other DIS processes
(see Table 1), while the results obtained in Refs. \cite{EK2,AR1}
without HT corrections are in better agreement with the LEP
measurements.
The estimated
effects of the HT contributions \cite{BBK},
 were also included
in the studies  of Ref. \cite{ANR}
aimed at a careful analysis  of the SMC and E142 data, using
the detailed parametrization of $g_1(x)$ at different values
of $x$ and fixed values of $Q^2$. In the process of another
interesting work  \cite{VShN} the authors  extracted
the values of the HT contributions directly from
 the experimental data. The results obtained are in qualitative
agreement with the outcomes of the QCD sum rules analysis \cite{BBK}.
 It is also worth while to mention that
the authors of  Ref. \cite{BI} advocate the conclusion that
the HT contributions to the Ellis-Jaffe SR might be even more
important than the ones to the polarized Bjorken SR.
All these examples demonstrate that
the problem of better   understanding
of the effects of the HT terms is more than a
pure theoretical one.
The  detailed information about the values of the HT corrections
will be even more important for the study of the theoretical
predictions for moments of the $g_2(x)$ structure function of the
polarized DIS. Notice that the experimental measurements of this
structure function are already pushed ahead \cite{SMC2}. However,
in view of the absence of  theoretical information
about the behaviour of $g_2(x)$ in the region of small $x$,
these measurements did not  yet allow us to check the validity
of the Burkhardt-Cottingham sum rule
\begin{equation}
\int_0^1 \, g_2(x,Q^2)dx=0
\label{bcsr}
\end{equation}
which  contrary to the claim of Ref. \cite{Neerven2}
does not receive perturbative QCD corrections \cite{Alt2}.

To conclude the discussions of this section we would like
to emphasize that the experimental and theoretical studies
of the polarized DIS functions are really important. These
offer new understanding of the
details of the nucleon structure. Clearly the studies of
these problems should be continued, and we believe that
the time and money necessary for the continuation of the
future experimental investigations in this field will  not
be wasted.

\section{Fragmentation Functions and
Unpolarized Deep- Inelastic Scattering}
There is a close analogy between the considerations of the
theoretical behaviour of the fragmentation functions (FFs)
$D(z,\mu^2)$ of the process $e^+e^-\rightarrow hX$, discussed
at the Symposium in the talk of Mele \cite{Mele} and those
of the
SFs $F_i(x,Q^2)$ of the DIS processes. Indeed, as in the case
of the SFs, the high-energy behaviour of the FFs is governed by the
Altarelli-Parisi equation, which in the non-singlet case has the
following form;
\begin{equation}
\frac{dD^{NS}(x,\mu)}{d\ln(\mu^2)}=\frac{\alpha_s(\mu^2)}{\pi}
\int_x^{1} \, \frac{dy}{y}D^{NS}(y,\mu)V_{NS}(x/y,\mu) ,
\label{ap}
\end{equation}
where the Mellin moments from the splitting function $V_{NS}(x)$
 determine the corresponding anomalous dimension function:
\begin{equation}
\int_0^{1} \, x^{n-1}V_{NS}(x)dx =\gamma_
{D}^{(n)}(\alpha_s)=
\sum_{i\geq 0} \, \gamma_{i}^{(n)}\bigg(\frac{\alpha_s}{\pi}\bigg)^{i+1}.
\label{anom2}
\end{equation}
In the talk of Mele \cite{Mele}, which was based on the detailed
discussion of Ref. \cite{NasWeb}, it was stressed that the
calculations of the NLO perturbative QCD approximation of the
FFs $D_i(x,\mu)$ are not yet completed. The  important
problem still remaining
is the NLO calculation of the longitudinal FF, which
can provide a new method of measuring $\alpha_s$ \cite{Mele},
\cite{NasWeb}.

 Contrary to the case of FFs, the behaviour of the DIS SFs
and in particular of the non-singlet ones is known at the NNLO.
At this order of perturbation theory the expression for the
SF $F_2(x,Q^2)$ is known from the results of Ref. \cite{Van3}, while
the NNLO corrections to the SF $xF_3(x,Q^2)$ of the deep-inelastic
neutrino scattering were calculated in Ref. \cite{NZ1}.
It should be stressed that it is possible to check the results
of the complicated calculations of Ref. \cite{Van3} by comparing
them with
the ones  obtained in Ref. \cite{LV2} for the Mellin
moments:
\begin{equation}
M_{NS}^{(n)}(Q^2)=\int_0^1 \, x^{n-1}F_i^{NS}(x,Q^2)dx.
\label{mellin}
\end{equation}
These moments obey the RG equation with the anomalous- dimension
term. Its solution has the following form:
\begin{equation}
\frac{M_{NS}^{(n)}(Q^2)}{M_{NS}^{(n)}(Q_0^2)}=
exp\bigg[\int_{\alpha_s(Q_0^2)}^{\alpha_s(Q^2)} \,
\frac{\gamma_{F}^{(n)}(x)}{\beta(x)}dx\bigg]
\frac{C_{NS}^{(n)}(\alpha_s(Q^2))}{C_{NS}^{(n)}(\alpha_s(Q_0^2))} .
\label{momrg}
\end{equation}
In the case of $n=2,4,6,8,10$, the NNLO corrections to the
coefficient functions of the non-singlet moments of $F_2(x,Q^2)$
were calculated analytically  \cite{LV2}. It is pleasant to
stress that these results   coincide with the ones
obtained from the results of Ref. \cite{Van3} after taking the
corresponding Mellin moments. The possibility to compare the
cumbersome
results of Refs. \cite{LV2},  \cite{Van3} obtained by means
of different calculational methods demonstrates the attractive
feature of the exact analytical calculations.

The results of Refs. \cite{LV2}, \cite{Van3}
were taken into account in the
non-singlet fits of the BCDMS data in Ref. \cite{Gonzalo}.
The results of the fits are presented in Table 4 taken from
Ref. \cite{Gonzalo}.

%
%
%
%
\begin{tabular}{|c|c||c|c|c|c||} \hline
               Deuterium           &
                                    &
                  $\Lambda_{\ms}$ (MeV) &
                  $\kappa$ (MeV)      &
                  $\chi^2$($F_2$)/dof  &
                  $\chi^2$($R$)/dof     \\
\hline\hline
%
%
                    $F_2$   &
                      LO     &
             $ 182 \pm 32 $   &
                               &
                   $63.5/65$    &
                                \\
                                 &
                     NLO          &
             $ 182 \pm 30 $        &
                                    &
                   $62.7/65$         &
                                     \\
                            &
                  SI(N)      &
             $ 159 \pm 25 $   &
                               &
                   $62.4/65$     &
                                 \\
                                 &
                    NNLO          &
             $ 168 \pm 27 $        &
                                    &
                   $62.5/65$         &
                                     \\
                            &
                     SI(NN)  &
            $ 164 \pm 26 $    &
                               &
               $62.4/65$        &
                                 \\
\hline
%
%
             $F_2$ and $R$   &
                      LO      &
             $ 223 \pm 35 $    &
                                &
                   $65.1/65$       &
                   $91.5/43$        \\
                                 &
                       NLO        &
             $ 235 \pm 34 $        &
                                    &
                   $65.5/65$           &
                   $90.9/43$           \\


                            &
                      SI(N)  &
             $ 238 \pm 28 $   &
                               &
                 $70.8/65$        &
                 $66.3/43$         \\
                                 &
                      NNLO        &
             $ 218 \pm 27 $        &
                                    &
                   $65.5/65$           &
                   $79.9/43$           \\
\hline
%
%
       $F_2$ and $R$             &
                          LO      &
             $181\pm31$       &
                 $207\pm13$    &
                      $63.5/65$    &
                        $41.4/43$    \\

        (Including twist-4)      &
                   NLO            &
             $180\pm29$       &
                 $198\pm14$    &
                      $62.7/65$    &
                        $41.5/43$    \\
                            &
                      SI(N)   &
             $157\pm25$       &
                 $184\pm17$    &
                      $62.4/65$    &
                        $41.4/43$    \\
                                 &
                   NNLO           &
             $159\pm21$       &
                 $200\pm12$    &
                      $62.8/65$    &
                        $44.6/43$    \\
\hline
\hline
                  Proton            &
                                    &
                                     &
                                     &
                                      &
                                       \\
\hline\hline
%
%
                    $F_2$   &
                      LO     &
              $171 \pm 27$    &
                               &
                   $51.2/60$    &
                                \\
                                 &
                     NLO          &
             $ 175 \pm 26 $        &
                                    &
                   $47.6/60$         &
                                     \\
                            &
                      SI(N)  &
             $ 159 \pm 23 $   &
                               &
                   $46.0/60$     &
                                 \\
                                 &
                    NNLO          &
             $ 168 \pm 25 $        &
                                    &
                   $46.4/60$         &
                                     \\
                            &
                     SI(NN)  &
             $ 169 \pm 25 $   &
                               &
                   $45.8/60$    &
                                 \\
\hline
%
%
             $F_2$ and $R$   &
                      LO      &
             $ 200 \pm 30 $    &
                                &
                   $52.8/60$      &
                   $92.3/43$        \\
                                 &
                       NLO        &
             $ 222 \pm 29 $        &
                                    &
                   $50.0/60$           &
                   $82.8/43$           \\


                            &
                      SI(N)  &
             $ 231 \pm 25 $   &
                               &
                 $54.5/60$        &
                 $59.2/43$         \\
                                 &
                      NNLO        &
             $ 209 \pm 23 $        &
                                    &
                   $48.8/60$           &
                   $79.7/43$           \\
\hline
%
%
       $F_2$ and $R$             &
                          LO      &
              $ 168 \pm 27$          &
               $  205 \pm 14$         &
                $    52.0/60$          &
                      $ 43.0/43$        \\
        (Including twist-4)      &
                   NLO            &
              $ 175 \pm 26$          &
               $  195 \pm 14$         &
                   $ 48.2/60 $         &
                    $   43.5/43 $       \\
                                &
                      SI(N)       &
              $ 158 \pm 23 $         &
               $  179 \pm 17 $        &
                 $   46.6/60  $        &
                    $   42.7/43 $       \\
                                 &
                   NNLO           &
             $  158 \pm 22 $         &
               $  189 \pm 14 $        &
                $    46.5/60 $         &
                     $  47.8/43 $       \\
\hline
\end{tabular}
\baselineskip=11.0pt
{\small      \noindent
{\bf Table 4.} The results of the NS fits of the BCDMS data for $F_2$
and the SLAC data for $R=F_L/2xF_1$. The SI(N) and SI(NN)
lines indicate the outcomes of the scheme-invariant analysis at
the NLO and NNLO respectively. The fitted values of
$\Lambda_{\overline{MS}}$ are
 normalized to $f=4$ numbers of flavours.}
\newpage
The analysis of Ref. \cite{Gonzalo} was made with the help of the
method of the reconstruction of the SF from their moments using
the expansion of the SF over the Jacobi polynomials \cite{Jac1},
\cite{Jac2}:
\begin{equation}
F_i^{N_{max}}=x^{\alpha}(1-x)^{\beta}\sum_{n=o}^{N_{max}} \,
\Theta_n^{\alpha,\beta}(x) \sum_{j=0}^{n} \,
c_j^{(n)}(\alpha,\beta) M^{(j+2)}(Q^2)
\label{jacobi}
\end{equation}
where $c_j^{(n)}(\alpha,\beta)$ are the coefficients that
are expressed through $\Gamma$-functions and the parameters
$\alpha,\beta$ can be chosen such as to achieve the fastest
convergence of the series in the r.h.s. of
Eq. (\ref{jacobi}). It should be stressed that in order to
perform the fit of the data at the NNLO self-consistently
it is necessary to take into account the NNLO corrections to the
corresponding anomalous dimension functions. For the
 non-singlet moments $M_{NS}^{(n)}$ of the SF $F_2$, they
were recently analytically calculated in the case of $n=2,4,6,8$
\cite{LRV2}. For $n=3,5,7$ the NNLO coefficients of the
$\gamma^{NS}_F$-function were estimated  \cite{Gonzalo}
using the smooth interpolation
of the results of Ref. \cite{LRV2} to the case of odd moments.

Let us make several comments on the physical results presented
in Table 4. One can see that the inclusion directly in the
$\overline{MS}$ scheme of the NNLO corrections
in the analysis of the experimental data
 slightly decreases
by over 20 MeV
the NLO values of the parameter $\Lambda_{\overline{MS}}^{(4)}$.
Therefore, taking into account these effects cannot remove
the slight difference existing at present
between the central
values of $\alpha_s$ extracted from the
previous analysis of DIS
data (see Table 1) and from the LEP measurements. The most important
phenomenological outcome of the corresponding NNLO analysis
is the observation of the minimization of the difference between
the results obtained within the framework of the $\overline{MS}$
scheme and the scheme-invariant approach. This observation is
in agreement with the results of the study
 of the higher-order QCD corrections to $R(s)$ and to the
Gross-Llewellyn Smith SR within different approaches of
fixing the scheme-dependence ambiguities \cite{ChKL}, \cite{ChK}.

The progress
discussed above  in the detailed consideration of
different effects of perturbative QCD for the SFs and their moments
in the kinematical region $Q^2\rightarrow\infty$, $x$ fixed, should
not shadow down the continuous interest in the description of
the dynamics of the behaviour of the SFs in the limit
$x\rightarrow 0$, $Q^2$ fixed. The behaviour of the singlet
SFs in this limit is governed by the Balitsky-Fadin-Kuraev-Lipatov
(BFKL) equation. The partonic picture of the evolution of the
SF $F_2(x,Q^2)$ both for increasing $Q^2$ and for decreasing $x$ was
discussed at the Symposium by Kirschner \cite{Kir}.
It should be stressed that contrary to the case of
 polarized DIS,
theory is going ahead of the experiment in the investigations
of the physics typical to the
low $x$ limit. Indeed, the results of the
recent experimental measurements at HERA at values of $x$ down to
$10^{-4}$, presented in the detailed talk of Obrock \cite{Obrock},
demonstrate the fast rise of $F_2(x,Q^2)$ at low $x$-values. This
effect can be associated with the prediction, which follows from
the BFKL considerations, namely
\begin{equation}
F_2(x,Q^2) \approx \bigg(\frac{1}{x}\bigg)^{\omega_0}
\label{BFKL}
\end{equation}
where $\omega_0=(g^2/2\pi^2)N2\ln2$ and $N$ is the number of colours.
Another preliminary experimental result of HERA, namely the counting
of the number of the events typical of the forward jets in the
data of the H1 collaboration, is presented in Table 5, taken from the
report of Obrock \cite{Obrock}.
\begin{center}
\begin{tabular}{|c|c|c|c|} \hline
$x$ & data         & AP ($Q^2\rightarrow\infty$)     & BFKL
($x\rightarrow 0$)\\ \hline
$2\times 10^{-4}-1\times 10^{-3}$ & $85\pm9\pm17$
 & 37 &  75       \\ \hline
$1\times 10^{-3}-2\times 10^{-3}$ & $43\pm 7\pm9$
 & 32   &  36       \\  \hline
\end{tabular}
\end{center}
\baselineskip=11.0pt
{\small \noindent
{\bf Table 5.}
Comparison of the number of  events seen in the preliminary
data of H1 with theoretical expectations.}

One can conclude that this analysis is very promising to
understand whether HERA results are sensitive to the BFKL
evolution.

Apart from the physics at low $x$ the experimental program of HERA is
also includes the study of  ``classical'' problems. Indeed, as
was emphasized by Obrock \cite{Obrock}, it is possible to measure
the value of $\alpha_s$ at HERA using the process of the gluon jet
emission in the 2+1 outcoming jets. The preliminary results of these
measurements are compared with the LEP results at Fig. 8 taken
from the talk of Obrock \cite{Obrock}.

The preliminary result with all estimated uncertainties reads
\cite{Obrock} $\alpha_s(M_z)=0.121 \pm 0.003 \pm 0.003
^{+0.006}_{-0.007}\pm 0.005 \pm 0.006$. It is obvious that
more statistics are necessary in
order to get a more precise result.

The further accumulation of  experimental data is
also important for the more detailed studies of the physics of
low $x$. The theoretical investigations of the effects typical of
this region are also in the process of fast development. The
interesting possibility that the physics of low $x$ can be identical
to the physics of two-dimensional exactly solvable models was discussed
in detail by Kirschner \cite{Kir}. In one of the recent works it
was even shown that the low- $x$ limit can be described by the
generalization of the Heisenberg XXX model, namely by a periodic lattice
spin- 0 model \cite{FK}.

In conclusion of this section, let us return from the world of low $x$ to
the world of large $Q^2$, namely from HERA to LEP. In the talk of
F\"urstenau \cite{Fur} the new result of the extraction  of
$\alpha_s(M_Z)$ from the LEP data for FFs was presented. The obtained
preliminary value $\alpha_s(M_Z)=0.118 \pm 0.005$ is in agreement with
the world average value of $\alpha_s$ presented by Bethke \cite{Bethke}.
As was already mentioned, the theoretical formalism of the description
of the behaviour of the FFs is very similar to the one commonly used
in the analysis of the SFs. This fact gives us the idea that we can
expect soon the appearance of new, more detailed, results of the
determination of $\alpha_s$ from the LEP experimental data for FFs
based on the
generalization of the methods of the DIS analysis
 and their application to the physics of FFs.

\section{The Connections Between Physical Quantities}
In all previous discussions we  considered the $\overline{MS}$
scheme as the reference   method for fixing the scheme-dependence
ambiguities of the perturbative QCD predictions. However, there are
also other approaches to treat these ambiguities in the
higher orders of perturbation theory. The appearance of several
complete NNLO QCD results \cite{our}, \cite{ss}, \cite{BNP},
\cite{LTV}, \cite{LV} makes it possible to apply   these different
prescriptions in practice and to study the related phenomenological
and theoretical outcomes of these analyses. The definite steps
in the direction of  detailed studies of the results of
applications of the ECH prescription \cite{ECH} and of the
PMS method \cite{PMS} have already been done  \cite{ChKL},
\cite{MatSt},  \cite{ChK}.

Some other aspects of these
studies were discussed at the Symposium in the talk
by Brodsky \cite{Brodsky},
which was based on some recent work
 \cite{BLu}. In this work the ECH method (or the so-called
scheme-invariant perturbation theory \cite{SIP}) was used to
express the NNLO perturbative QCD predictions for one observable
quantity through another one. The considered physical quantities
were $R(s)$, $R_{\tau}$ and the DIS SRs. At the second step of their
considerations,
in order to get rid of the dependence on the number of flavours
$f$ in the resulting ``commensurate scale relations'',
the authors of Ref. \cite{BLu} applied the
BLM prescription \cite{BLM} and the variant of its NNLO generalization
\cite{GrK}.
Several
comments on the typical features of the methods used in the process
of the derivation of these relations, which provide the additional
possibilities for testing perturbative QCD predictions, are now in order.

First, the application of the ECH prescription does not eliminate
the ambiguities typical of the perturbative expansions. Indeed,
instead of solving the guess about the ``basic phenomenological
scheme'' for the definition of $\alpha_s$ one should now write
down the agreement about the basic physical quantity, which
should be used as the reference in all corresponding
perturbative expansions. The authors of Ref. \cite{BLu} are
proposing for
 this role the effective quark potential. However, in
view of the absence of  information about
 the NNLO perturbative
corrections to this quantity, it is impossible at present to
check the  advantages of this choice self-consistently.

Secondly, one should be careful in applying the BLM prescription
in the high orders of perturbation theory. For example, the
comparison of the results of the
application of this procedure to
the analysis of the effects of the perturbative QCD corrections
to the $\rho$ parameter \cite{Voloshin} with the results of
the exact calculations \cite{Avdeev} demonstrates that the BLM
approach is missing the  numerical
contribution most important in this case,
 which comes from the three-loop diagram with the triangle
fermion loop insertions. Quite fortunately, in the case of
$R_{\tau}$ and of the
polarized Bjorken sum rule the light-by-light-type
graphs are absent, while the contribution of these graphs to
$R(s)$ and to the Gross-Llewellyn Smith SR is small. However,
even in this case the applications of the main relation derived in
Ref. \cite{BLu}, namely
\begin{equation}
\hat\alpha_{g_1}(Q)=\hat\alpha_R(Q^*)-\hat\alpha_R^2(Q^{**})
+\hat\alpha_R^3(Q^{***})+...,
\label{crel}
\end{equation}
where $\hat\alpha=(3C_F/4\pi)\alpha$ and $\alpha_{g1}$ and
$\alpha_R$ are the effective charges of the polarized Bjorken
SR and $R(s)$, the BLM method is  shadowing the effects of new physics,
which obviously manifest themselves in the $\overline{MS}$
scheme.

Indeed, the simplicity of the relation of Eq. (\ref{crel})
is connected  with the Crewther quark-parton relation
\cite{Crew} between three fundamental quantities, namely
the anomalous constant $S$, associated with the amplitude
of the $\pi^0\rightarrow\gamma\gamma$ decay,
the polarized Bjorken SR BjpSR
 and the Adler $D$-function in the annihilation
channel. The derivation of this relation relied on the conformal
and the chiral invariance. However, it was shown  \cite{BrKat}
that
the na\'ive Crewther relation, at the NLO and NNLO of perturbation
theory, receives additional radiative
corrections, which are proportional to the $\beta$-function,
or more definitely to the factor $\beta(a)/a$:
\begin{equation}
\Delta_S=BjpSR\times D-1=\frac{\beta(a)}{a}\bigg[
S_1C_Fa+\bigg(S_2T_Ff+S_AC_A+S_FC_F\bigg)C_Fa^2\bigg]+O(a^4)
\label{dev}
\end{equation}
where  $C_F=4/3$, $C_A=3$, $T_F=1/2$ in the case of QCD
and $S_1=-21/2+12\zeta(3)$, $S_2=326/3-(304/3)\zeta(3)$,
$S_A=-629/2+(884/3)\zeta(3)$ and $S_F=397/6+136\zeta(3)-
240\zeta(5)$ are the analytical numbers independent of
 the structure of the gauge group.
It is now
possible  to understand that Eq. (\ref{crel})
can be obtained from Eq. (\ref{dev}) in the conformal invariant
limit, namely after the nullification of the
factor $\Delta_S$. However, the applications of the BLM approach
presented in
Ref.\cite{BLu} do not allow us
to understand the origine of
the appearence of this factor in the fixed schemes. Work
on the study of this fact is now in progress \cite{GA1}.
The puzzle,
discovered in Ref. \cite{BrKat}, of the cancellations
of the $C_F^{N}$ factors in the $N$-th order approximation for
$\Delta_S$ is already understood \cite{GA2}. This cancellation
is the consequence of the Adler-Bardeen theorem.
The rigorous
theoretical explanation of the remaining ``wonders'' from
the ``seven'' ones observed in Ref. \cite{BrKat} is on the agenda.

We hope that the possible outcomes of these studies will have
both theoretical and phenomenological consequences, which might
be important for the more detailed understanding of the
experimental numbers obtained
for the polarized Bjorken SR and for
the Gross-Llewellyn Smith SR, extracted from the data
of the CCFR collaboration \cite{CCFR} in Refs. \cite{CCFR1},
\cite{Sid}.

\section{Conclusion}
Summarizing the discussions of the current status of QCD,
presented at this Symposium, it is necessary to stress that
in general this theory is in  good condition. However,
a lot of important problems are still waiting for a solution.
Among them are
\begin{enumerate}
\item the calculations of the NLO corrections to the 4-jets
characteristics;
\item
more rigorous and precise determinations of the values
of light and heavy quark masses;
\item
the necessity to  understand  the origin of the
certain discrepancies between the values of $\alpha_s$ extracted
from different processes.
\item
Clearly, the desire to understand this problem is supporting new
experimental measurements and new more precise calculations, say the
above-
mentioned  calculations of the higher-order perturbative
QCD corrections to the cross sections of the $e^+e^-\rightarrow
4$ jets and $\overline{p}p\rightarrow 3$ jets processes.
\item
There are a number of  calculational problems, which are
related to the recent development of the new rigorous theory
of the study of the heavy quarkonium annihilation rates \cite{Eric2},
which was discussed   at the Symposium in the
talk by Braaten \cite{Eric}. This theory allows to calculate the
related characteristics
from the first principles, the only input being the heavy-quark mass
and the QCD coupling constant. Concrete applications of this formalism
are on the agenda.
\item
The next important problem is related to the necessity of
 more detailed studies
of the effects of the HT corrections in the polarized DIS.
The future works can provide a more rigorous control of this type
of the theoretical uncertainties in the description of the
corresponding experimental data. This problem is of particular
interest in view of the existence of  programs of future
measurements of the polarized DIS SFs at SLAC and CERN.
\item
In a future analysis of the new data on both polarized
and non-polarized DIS the problem of the parametrization
of the behaviour of the SFs at low $x$ should be considered
very carefully. Note that this limit is not totally described
by the perturbation theory. The possibilities provided by
the non-perturbative methods, and in particular by the mathematical
methods typical of the exactly solvable models, should be developed
to the new level.
\item
There are still a
number of problems that should be understood on the new level
within the framework of the perturbation theory.
For example different points of view on
different methods of treating the scheme-dependence
problem still do not allow us
to write down the convention about
the best way of controlling  the theoretical uncertainties
in the analysis of the effects of the
higher-order  perturbative QCD corrections.
\item
Even more important can  be the breakthrough in the understanding
of the theoretical structure of the perturbative series in
QCD and gauge models.
\item
It is important to find some  new methods of testing the predictions
of the perturbative QCD. We believe that the future, more
detailed studies of the  Crewther relation
\cite{Crew} and its different generalizations \cite{BrKat},
\cite{BLu} might give us the opportunity of comparing theoretical
and experimental results
for the DIS SRs with the ones for the annihilation
processes on the new level.
\end{enumerate}

To conclude, let us hope that the era of  perturbative
QCD is not finished and that we can expect the appearance
of  new interesting and important results.

\section{Acknowledgements}
We would like to thank B. L. Ward for his invitation to
this very productive Symposium and for the warm hospitality
in Tennessee. We are grateful to J. Ellis, M. Karliner and
F. Le Diberder for providing us
the possibility to present
the results of their works before  publication.

\end{document}